\documentclass[epj]{svjour}
\usepackage{graphics}
\usepackage{caption}
\usepackage{subcaption}
\RequirePackage[T1]{fontenc}
\usepackage{mathtools}

\usepackage{float}
\RequirePackage{graphicx}
\RequirePackage{mathptmx}      
\RequirePackage{flushend}
\RequirePackage[numbers,sort&compress]{natbib}
\RequirePackage[colorlinks,citecolor=blue,urlcolor=blue,linkcolor=blue]{hyperref}

\usepackage[numbers]{natbib}
\usepackage{rotating}
\usepackage{lscape}
\usepackage{graphicx}
\usepackage{braket}
\usepackage{multirow}
\usepackage{amsmath}
\begin{document}
\title{Probing Kaons as Light-Strange Tetraquarks through Spectral and Decay Dynamics}
\author{Chetan Lodha\inst{1},
	 \and  Ajay Kumar Rai\inst{2},}                     

%
\institute{Department of Physics, Sardar Vallabhbhai National Institute of Technology, Surat, Gujarat-395007, India, \\ \email{iamchetanlodha@gmail.com }\\ \email{raiajayk@gmail.com}}

\mail{iamchetanlodha@gmail.com}
\date{Received: date / Revised version: date}
%
\abstract{
	Motivated by the discovery of several kaon-like states observed at BESIII and LHCb, this study explores S-wave and P-wave tetraquark states with quark contents \( ss\bar{s}\bar{q} \) and \( sq\bar{q}\bar{q} \) using potential-based phenomenology. The tetraquarks are modeled as diquark-antidiquark systems in antitriplet-triplet and sextet-antisextet color configurations. The mass spectra are calculated using the Cornell potential, while decay properties are analyzed through Fierz rearrangement. A comparison with experimentally observed states is provided to enhance our understanding of light-light tetraquark systems with non-homogeneous quark compositions.
\PACS{
      {PACS-key}{discribing text of that key}   \and
      {PACS-key}{discribing text of that key}
     } 
} 

\maketitle
\section{Introduction}


Hadron spectroscopy provides crucial insights into the internal structure of hadrons and the dynamics of strong interactions governing their behavior. Conventional hadrons, such as baryons and mesons, were initially described by the quark model \cite{Gell-Mann:1964ewy,Berwein:2024ztx} and have since been extensively studied and refined through both experimental observations and theoretical advancements. Experimentally, cutting-edge facilities like Belle \cite{Berwein:2024ztx}, PANDA \cite{Gotzen:2024agc,Belias:2023lkk}, J-PARC \cite{KOTO:2024zbl,Shiomi:2022rgi}, LHCb \cite{Gandini:2025eii}, and others have made significant efforts to identify resonances that could serve as potential candidates for these hadrons. On the theoretical front, frameworks such as the MIT Bag Model \cite{Chodos:1974je}, QCD Sum Rules \cite{Agaev:2024wvp}, Lattice QCD \cite{Hernandez-Pinto:2024kwg}, Flux Tube Model \cite{Jakhad:2024fgt,Jakhad:2024fin}, Regge Phenomenology \cite{Oudichhya:2023lva}, Quark Meson Coupling Model \cite{Mondal:2023iwe}, Hypercentral Constituent Quark Model (HCQM) \cite{Menapara:2024wpb,Menapara:2024uea} and Potential Phenomenology \cite{Patel:2024cng}, among others, have been employed to analyze these resonances and uncover the underlying physics. 

The study of exotic hadrons has become a transformative area in hadronic physics, engaging both experimentalists and theorists. These unconventional states deviate from the traditional quark model, offering insights into the strong interaction and expanding the boundaries of particle physics. Since their proposal in the late 1970s \cite{Jaffe:1976ig}, exotic hadrons have been explored through methods like QCD Sum Rules \cite{Agaev:2024wvp,Yang:2024okq}, Potential Phenomenology \cite{Tiwari:2021iqu,Tiwari:2021tmz,Mistry:2024zna}, Lattice QCD \cite{Radhakrishnan:2024ihu}, and the Flux Tube Model \cite{Pal:2024yqe}, enriching our understanding of quantum chromodynamics (QCD). These exotic hadrons can be explained as

\begin{itemize}
	\item Hybrid Meson : A bound state including quark-antiquark-gluon $(q\bar{q}g)$ configuration \cite{Akbar:2024jda}.
	\item Hadronic Molecule : A state of two hadrons bound together by strong interaction, mimicking traditional molecule-like state \cite{Rathaud:2019tit,Rai:2006wm}. 
	\item Compact Tetraquark : A bound state of diquark-antidiquark $([qq][\bar{q}\bar{q}])$ configuration, mimicking meson-like state \cite{Tiwari:2021iqu,Tiwari:2021tmz,Mistry:2024zna}.
	\item Compact Pentaquark : A bound state of diquark-diquark-antiquark $([qq][qq]\bar{q})$ configuration, mimicking baryon-like state \cite{Yan:2023iie}.  
	\item Glueball : A bound state made-up entirely of gluons \cite{Li:2024fko}.
\end{itemize}

These states challenge traditional paradigms and require novel theoretical approaches to understand their formation, decay, and interactions, providing a window into the non-perturbative regime of QCD. The experimental discovery of exotic hadrons began with a robust tetraquark candidate observed in 2003 \cite{Belle:2003nnu}. Since then, numerous candidates have been identified in experiments at Belle \cite{Belle:2021kub}, BESIII \cite{BESIII:2022vxd}, and LHCb \cite{LHCb:2024xyx,LHCb:2023evz}, contributing critical data on production, decay, and kinematic properties. These findings refine theoretical models and validate the existence of configurations beyond the Standard Model.

Kaons are among the most elementary bound states observed in particle physics. Despite their simple construction, kaons have advanced our understanding of intricate and fundamental phenomena such as CP violation, state mixing, and weak interactions. These mesons, with their unique quark compositions, have long been a focus of experimental and theoretical research, yielding invaluable insights into the dynamics of the Standard Model and beyond. Over the decades, significant progress in kaon physics has been driven by cutting-edge experiments, namely the NA48 and NA62 experiments at CERN, the KOTO experiment at J-PARC, the LHCb experiment and the E787 experiment at Brookhaven National Laboratory. These facilities have provided critical data on CP violation, flavor physics involving kaons and rare decay processes and have significantly enhanced our understanding of kaons. Remarkably, some kaon resonances listed by the Particle Data Group (PDG) \cite{ParticleDataGroup:2024cfk} show intriguing compatibility with theoretical descriptions of tetraquark states. These tetraquark include configurations with quark content such as \( ss\bar{s}\bar{q} \) and \( sq\bar{q}\bar{q} \). This observation suggests that certain kaon resonances could potentially be interpreted as tetraquark candidates. The present work investigates these resonances to assess their viability as exotic states, with the aim of expanding our understanding of hadronic structures beyond traditional quark models. 

The current paper is structured as follows: Followed by a brief introduction in Section I, Section II illustrates the theoretical framework for mass spectroscopy. Section III is dedicated to explaining the mass spectra of tetraquarks. Section IV presents the decay properties of tetraquarks. Section V presents the regge plots of tetraquark. In Section VI, we present the results and discussion and finally conclude in Section VII.

\section{Theoretical Framework}
\label{sec:1}
Inspired by studies \cite{Lodha:2024yfn,Lodha:2024bwn,Lodha:2024qby}, a phenomenology-inspired potential that elucidate constituent quark interactions are employed in a semi-relativistic as well as the non-relativistic framework. A compact tetraquark is formed when a pair of diquark and anti-diquark are held together by color force. Since the spin-1 $ss$, $qq$ and $sq$ diquarks / anti-diquarks have a mass nearing 1 GeV, putting them at a cusp of both semi-relativistic and non-relativistic realm. Hence, both semi-relativistic and non-relativistic frameworks are employed to understand the mass spectra of kaonic tetraquark. As illustrated with in-depth detail in ref. \cite{Eichten:1979ms},
a static potential-inspired non-relativistic approach is used for a tetraquark system as a reasonable approximation. The binding energy of each specific state is determined using the solution of the time-independent radial Schrodinger equation. To encompass a two-body problem in the central potential model, the idea of using the center of mass frame is very effective. Using spherical harmonics, the angular and radial terms of the time-independent Schrodinger wave function can be isolated. The fundamental two-body semi-relativistic Hamiltonian with relativistic correction to the kinetic energy is given by $H_{SR}$. The Hamiltonion for  non-relativistic framework in the center of mass frame is  given by $H_{NR}$,

\begin{equation}
	H_{SR} = \sum_{i=1}^2 \sqrt{p^{2}_{i}+M^{2}_{i}} + V^{(0)} (r) + V_{SD}(r),
\end{equation}

\begin{equation}
	H_{NR} = \sum_{i=1}^2 (M_{i} + \frac{p_{i}^{2}}{2M_{i}}) + V^{(0)} (r) + V_{SD}(r).
\end{equation}
where $M_{i}$, $p_{i}$ and V (r) are the constituent mass, relative momentum of the system, and interaction potential, respectively. The kinetic energy term of the semi-relativistic Hamiltonian is expanded up to $\mathcal{O}(p^{10})$ as,

\begin{equation} 
	\begin{split}
		K.E. = & \sum_{i=1}^2 \frac{p^{2}}{2}\bigr(\frac{1}{M_{i}}\bigr) - \frac{p^{4}}{8}\bigr(\frac{1}{M_{i}^{3}}\bigr) + \frac{p^{6}}{16}\bigr(\frac{1}{M_{i}^{5}}\bigr) \\ 
		&- \frac{5p^{8}}{128}\bigr(\frac{1}{M_{i}^{7}}\bigr) + \frac{7p^{10}}{256}\bigr(\frac{1}{M_{i}^{9}}\bigr) + \mathcal{O}(p^{12}).
	\end{split}
\end{equation}

In the expansion up to \(\mathcal{O}(p^{10})\), the \(\mathcal{O}(p^{4})\) and \(\mathcal{O}(p^{8})\) terms contribute negatively, whereas the \(\mathcal{O}(p^{6})\) and \(\mathcal{O}(p^{10})\) terms contribute positively, indicating potential cancellations among these terms. This dynamic interplay underscores the importance of the \(\mathcal{O}(p^{10})\) term in shaping the mass spectra of strange tetraquarks, highlighting its critical role in modulating their mass distribution. A detailed discussion of this expansion is presented in our previous study, Ref. \cite{Lodha:2024qby}.

While the zeroth-order Cornell-like potential is an extensively used phenomenological model for the study of heavy quarkonia spectroscopy, numerous other methods have also been employed to study hadronic mass spectra \cite{Eichten:1979ms,Quigg:1979vr,Godfrey:1985xj}. The Cornell-like potential used in this study incorporates two key components: a Coulombic term and a linear term. The Coulombic potential arises from a Lorentz vector exchange, specifically in the form of a one-gluon exchange between quarks, which mimics the short-range interaction similar to the Coulomb force in electrodynamics. This Coulombic term dominates at short distances between quarks, where the strong force behaves in a manner akin to the electromagnetic force. On the other hand, the linear potential is a result of Lorentz-scalar exchange, representing the mechanism of quark confinement. This component reflects the long-range behavior of the strong force, where the potential energy increases linearly as the distance between the quarks grows. This linear dependence effectively models the phenomenon of confinement, where quarks become increasingly bound together as they move further apart, preventing their isolation as free particles. While the precise mechanism of confinement in quantum chromodynamics (QCD) remains one of the fundamental open questions, the linear nature of the confinement potential is well-supported by  theoretical models. Lattice QCD simulations, which offer a numerical solution to QCD on a discretized spacetime lattice, further reinforce this linear relationship. Thus, the Cornell-like potential, combining both Coulombic and linear terms, provides a simplified yet powerful framework to describe the quark-quark interactions, balancing short-range and long-range dynamics in hadron physics. The zeroth-order Cornell-like potential is given by, 

\begin{equation}
	V^{(0)}_{C+L}(r) = \frac{k_{s}\alpha_{s}}{r} + br+ C,
\end{equation}

where $\alpha_{s}$, $k_{s}$, $b$, $C$ are the running coupling constant, color factor, string tension and scaling parameter, respectively. The color factor values for different states are $-\frac{4}{3}$ for the color-singlet state, $-\frac{2}{3}$ for the triplet-antitriplet state, and $\frac{1}{3}$ for the sextet state. Drawing inspiration from references \cite{Koma:2006si}, a relativistic mass correction has been introduced to the central potential to account for the effects of relativity on quark dynamics. This adjustment is crucial for refining the potential model, as it ensures a more accurate representation of the interactions between quarks, particularly when considering systems where relativistic effects cannot be neglected. Hence, the central potential finally manifests as,

\begin{equation}
	V^{(0)}(r) = V^{0}_{C+L}(r) + V^{1}(r) \biggl(\frac{1}{m_{1}} + \frac{1}{m_{2}} \biggl),
\end{equation}
where $m_{1}$ and $m_{2}$ are the constituent masses of constituent particles of the bound state, namely diquarks. Considering the fact that the non-perturbative form of the relativistic mass correction term is still unknown, the leading order of the perturbation theory is utilized, which yields

\begin{equation}
	V^{1}(r) = - \frac{C_{F}C_{A}}{4} \frac{\alpha_{s}^{2}}{r^{2}},
\end{equation}

where $C_{F}$ and $C_{A}$ are the Casimir charges of the fundamental and the adjoint representation, respectively \cite{Koma:2006si}. As demonstrated in \cite{Brambilla:1999xf}, this calculation aligns consistently with the one-loop contribution. The physics of \( \frac{1}{m} \) corrections to the QCD potential is insufficiently studied and underscores the necessity for further investigation in this area. In order to understand the splitting between the orbital and radial excitations of various states, spin-dependent interactions are crucial and have been incorporated perturbatively \cite{Lucha:1991vn}.

\subsection{Spin-dependent Interactions}

Inspired by the Breit-Fermi Hamiltonian for one gluon exchange, three spin-dependent interactions are introduced \cite{Lucha:1991vn}.
\begin{equation}
		V_{SD}(r)  = V_{T}(r) + V_{LS}(r) + V_{SS}(r). 
	\end{equation}
	\begin{subequations}
		\begin{equation}
		= \biggl( - \frac{k_{s}\alpha_{s}}{4} \frac{12\pi}{M_{\mathcal{D}}M_{\bar{\mathcal{D}}}}\frac{1}{r^{3}}\biggl) \; \biggl(-\frac{1}{3}(S_{1}\cdotp S_{2}) + \frac{(S_{1} \cdotp r) {(S_{2}\cdotp r)}}{r^{2}}\biggl) 
		\label{eqvt}
\end{equation}
\begin{equation}
		 + \biggl(-\frac{3\pi k_{s}\alpha_{s}}{2M_{\mathcal{D}}M_{\bar{\mathcal{D}}}}\frac{1}{r^{3}}  -  \frac{b}{2M_{\mathcal{D}}M_{\bar{\mathcal{D}}}}\frac{1}{r}   \biggl)(L\cdotp S) 
		 \label{eqls}
\end{equation}
\begin{equation}
		+ \biggl(- \frac{k_{s}\alpha_{s}}{3} \frac{8\pi}{M_{\mathcal{D}}M_{\bar{\mathcal{D}}}} \frac{\sigma}{\sqrt{\pi}}^{3} exp^{-\sigma^{2}r^{2}}\biggl) (S_{1}\cdotp S_{2})
		\label{eqss}
\end{equation} 
\end{subequations}

where the masses of diquark and antidiquark are represented by $M_{\mathcal{D}}$ and $M_{\bar{\mathcal{D}}}$. The solution of diagonal matrix elements of spin-$\frac{1}{2}$ and spin-1 particles determines the value of $(S_{1}\cdotp S_{2})$ \cite{Debastiani:2017msn}. $\sigma$ is a parameter that has been introduced as a replacement for the Dirac Delta function. 

Equation \ref{eqvt} in $V_{SD}(r)$ represents the tensor interaction potential, while equation \ref{eqls} corresponds to the spin-orbit interaction potential, and equation \ref{eqss} describes the spin-spin interaction potential. These spin-dependent interaction potentials have been discussed extensively in our previous work \cite{Lodha:2024qby}, with further elaboration provided in studies such as \cite{Tiwari:2021iqu,Debastiani:2017msn,Lundhammar:2020xvw}. The spin-orbit interaction potential, $V_{LS}$, and the tensor interaction potential, $V_{T}$, contribute to the fine structure of the states, affecting their energy levels by introducing small corrections. On the other hand, the spin-spin interaction potential, $V_{SS}$, is responsible for hyperfine splitting, which differentiates states that have the same spatial configuration but different total spin. These spin-dependent interactions are typically handled using first-order perturbation theory, where their matrix elements are treated as energy corrections to the overall potential. Each of these interaction potentials is formulated in terms of the static quark potential, $V(r)$, which describes the interaction between quarks as a function of their separation distance. For a more accurate description of tetraquark spectroscopy, the spin-spin interaction can be included directly in the zeroth-order potential, providing a reasonable approximation of the mass spectra. This approach captures the key contributions of spin-dependent forces in determining the fine and hyperfine structure of the tetraquark states. The $J^{P}$ values of different tetraquark states can be obtained by using the relations 	$P_{T} = (-1)^{L_{T}} $ \cite{Lucha:1995zv}, where $L_{T}$ is the total angular momentum, respectively. 

	\section{Spectroscopy}
	As previously discussed, when a diquark and an anti-diquark are bound together by the color force, they can form a compact tetraquark in a color-singlet configuration. This singlet state can arise in two distinct configurations, depending on the color representation of the diquark and anti-diquark involved. The most commonly studied configuration of a color-singlet tetraquark is formed by a diquark in the anti-triplet ($\bar{3}$) color representation and an anti-diquark in the triplet ($3$) configuration. In this case, the color force binds the two components, with the system's color factor being $k_s = -\frac{4}{3}$. The resulting singlet tetraquark is typically described by combining a spin-1 diquark and a spin-1 anti-diquark. The color representation of this configuration is denoted as $\ket{QQ|^{\bar{3}}\otimes|\bar{Q}\bar{Q}|^{{3}}} = \textbf{1}\oplus\textbf{8}$, indicating that the product of these two color states includes both a color-singlet (\textbf{1}) and an octet (\textbf{8}), but only the singlet contributes to the physical state. 	Alternatively, some studies propose a different compact tetraquark model where the diquark is in a sextet ($6$) configuration, and the anti-diquark is in an anti-sextet ($\bar{6}$) configuration. In this scenario, the two are still held together by the color force, but with a larger color factor of $k_s = -\frac{10}{3}$. Here, the singlet tetraquark is formed by combining a spin-0 diquark and a spin-0 anti-diquark. The color structure of this configuration is more complex, represented by $\ket{QQ|^{6}\otimes|\bar{Q}\bar{Q}|^{\bar{6}}} = \textbf{1}\oplus\textbf{8}\oplus\textbf{27}$, where, in addition to the singlet and octet, a higher-dimensional color state (\textbf{27}) emerges. These two configurations highlight different possible internal color structures within tetraquarks, where the anti-triplet-triplet combination is more commonly explored due to its simpler color dynamics, while the sextet-anti-sextet configuration offers an alternative but less frequently discussed model of tetraquark structure. The distinction in the color factor values and spin states emphasize the diverse ways in which the strong force can organize quarks within exotic hadrons like tetraquarks. The parameters $b,\alpha_{s},\sigma, M_{ss},M_{sq}$ and $M_{qq}$ for the present work are adopted from our previous work, ref. \cite{Lodha:2024yfn,Lodha:2024bwn,Lodha:2024qby}. The tetraquark mass spectra are calculated by :
	 \begin{subequations}
\begin{equation}
	M_{(ss\bar{s}\bar{q})} = M_{ss} + M_{\bar{s}\bar{q}} + E_{(ss\bar{s}\bar{q})} + \braket{V^{1}(r)}.
\end{equation} 
\begin{equation}
	M_{(sq\bar{q}\bar{q})} = M_{sq} + M_{\bar{q}\bar{q}} + E_{(sq\bar{q}\bar{q})} + \braket{V^{1}(r)}.
\end{equation} 
\end{subequations}
The mass of the calculated tetraquark is obtained with the mass contribution from the Cornell-like potential, relativistic term $\braket{V^{1}(r)}$. All spin-dependent contributions are calculated individually. By coupling total spin $S_{T}$ with orbital angular momentum $L_{T}$, $S_{T}\otimes L_{T}$, the color singlet state of the tetraquark is obtained.
\begin{equation}
	\ket{T_{4s}} = \ket{S_{d},S_{\bar{d}},S_{T},L_{T}}_{J_{T}},
\end{equation}
where $S_{d}$ and $S_{\bar{d}}$ are the spins of diquark and anti-diquark, respectively. Ref. \cite{Lodha:2024qby} explores the dynamics of possible states and their configurations, as well as the parameter sensitivity of the model in detail. The mass spectra of S wave and P wave $\bar{\textbf{3}}\otimes{\textbf{3}}$ and ${\textbf{6}}\otimes\bar{\textbf{6}}$ states are done tabulated in table \ref{Swavetriplet}. The two meson threshold for various states is tabulated in table \ref{twomesonthreshold}.

\begin{table*}
	\caption{S Wave SSSQ All units are in MeV.}
	\label{Swavetriplet}
	\begin{tabular*}{\textwidth}{@{\extracolsep{\fill}}llcccc|cccc@{}}
		\hline
		\multirow{3}{*}{State} &	\multirow{3}{*}{$J^{P}$} & \multicolumn{4}{c|}{$T_{sssq}$}  & \multicolumn{4}{c}{$T_{sqqq}$} \\

		\cline{3-10}
		& & \multicolumn{2}{c}{$\bar{3}\otimes{3}$}& \multicolumn{2}{c|}{$6\otimes\bar{6}$} & \multicolumn{2}{c}{$\bar{3}\otimes{3}$}  & \multicolumn{2}{c}{$6\otimes\bar{6}$} \\
		\cline{3-10}
		& & \multicolumn{1}{c}{M$_{SR}$}  & \multicolumn{1}{c}{M$_{NR}$} & \multicolumn{1}{c}{M$_{SR}$}  & \multicolumn{1}{c|}{M$_{NR}$} & \multicolumn{1}{c}{M$_{SR}$}  & \multicolumn{1}{c}{M$_{NR}$}& \multicolumn{1}{c}{M$_{SR}$}  & \multicolumn{1}{c}{M$_{NR}$} \\
		
		\hline
		$1^{1}S_{0}$ &\multirow{3}{*}{$0^{+}$} & 1688.83 & 1845.90 & 1161.27 & 899.72  & 1439.99 & 1239.35 & 1631.54 & 739.30 \\
		$2^{1}S_{0}$ & & 2705.68 & 2797.19 & 2932.67 & 3104.81 & 2470.83 & 2790.35 & 2595.88 & 2973.53 \\
		$3^{1}S_{0}$  && 3181.37 &3280.76 &3879.06 &4055.92 &2962.87&3091.70 &3749.25 &3932.74\\
		\hline
		$1^{3}S_{1}$ &\multirow{3}{*}{$1^{+}$} & 1871.25 & 1998.89 & - & - & 1653.02 & 1804.26 & - & - \\
		$2^{3}S_{1}$ & & 2754.48 & 2840.05 & - & - & 2535.26 & 2647.59 & - & - \\
		$3^{3}S_{1}$ & & 3214.45 &3306.98 &- &- & 3003.36&3122.71 &- &- \\
 	 	\hline
		$1^{5}S_{2}$ &\multirow{3}{*}{$2^{+}$} & 2338.73 & 2300.37 & - & - & 2070.76 & 2140.97 & - & - \\
		$2^{5}S_{2}$ & & 2852.10 & 2918.98 & - & - & 2658.23 & 2742.72 & - & - \\
 		$3^{5}S_{2}$ & & 3277.07 &3359.33 &- & -& 3083.88&3187.37 &- &- \\
		\hline
		$1^{1}P_{1}$ &\multirow{3}{*}{$1^{-}$} & 2635.17 & 2710.62 & 2706.09 & 2868.47 & 2415.16 & 2513.20 & 2561.05 & 2732.62 \\
		$2^{1}P_{1}$ & & 3091.39 & 3175.25 & 3692.30 & 3850.09 & 2876.64 & 2985.56 & 3552.21 & 3721.48 \\
 		$3^{1}P_{1}$ & & 3455.38 &3554.63 &4403.68 &4580.34 &3248.90 &3374.29 &4278.30 & 4463.72\\
 		\hline
		$1^{3}P_{0}$ &\multirow{3}{*}{$0^{-}$} & 2314.79 & 2390.20 & 1010.72 & 1151.97 & 2082.28 & 2180.37 & 841.42 & 988.89 \\
		$2^{3}P_{0}$ & & 2821.00 & 2905.99 & 2353.68 & 2501.65 & 2590.99 & 2701.27 & 2173.36 & 2327.59 \\
 		$3^{3}P_{0}$ & & 3206.10 &3304.82 &3204.07 &3376.79 &2982.91 &3109.17 &3034.28 &3212.69 \\
 \hline
		$1^{3}P_{1}$ &\multirow{3}{*}{$1^{-}$} & 2642.87 & 2715.40 & 2700.65 & 2842.13 & 2425.86 & 2520.79 & 2564.90 & 2712.98 \\
		$2^{3}P_{1}$ & & 3095.55 & 3177.21 & 3655.01 & 3801.86 & 2883.79 & 2989.50 & 3521.30 & 3676.27 \\
 		$3^{3}P_{1}$ & & 3458.16 & 3553.31 &4356.13 &4524.83 &3254.76 &3375.03 &4223.66 &4411.05 \\
 \hline
		$1^{3}P_{2}$ &\multirow{3}{*}{$2^{-}$} & 2746.45 & 2819.63 & 3291.55 & 3432.29 & 2531.40 & 2624.60 & 3157.82 & 3305.57 \\
		$2^{3}P_{2}$ & & 3184.20 & 3266.30 & 4114.87 & 4261.56 & 2976.82 & 3081.22 & 3993.93 & 4146.93 \\
 		$3^{3}P_{2}$ & &3540.91 & 3636.28 &4765.83 &4935.03 &3342.08 &3461.94 &4659.42 &4834.89 \\
 \hline
	
		$1^{5}P_{1}$ &\multirow{3}{*}{$1^{-}$} &  2321.96  & 2388.86 &- &- & 2091.60 & 2181.63 &- &- \\
		$2^{5}P_{1}$ & & 2824.87  & 2901.96 &- &-& 2598.23  & 2699.94 &-&- \\
 		$3^{5}P_{1}$ & & 3206.11 &3297.28 &- &- &2986.94 &3103.91 &- &- \\
 \hline
		$1^{5}P_{2}$ &\multirow{3}{*}{$2^{-}$} & 2732.50  & 2797.72 &- &- & 2522.01  & 2605.59 &-&- \\
		$2^{5}P_{2}$ & & 3168.42  & 3243.30 &- &-& 2965.75  & 3060.92 &-&- \\
 		$3^{5}P_{2}$& & 3521.48 &3610.14 &- &- &3326.73 &3437.96 &- &- \\
 \hline
		$1^{5}P_{3}$ &\multirow{3}{*}{$3^{-}$} & 2887.23  & 2951.52 &- &-& 2677.03  & 2762.21 &-&- \\
		$2^{5}P_{3}$ & & 3302.98  & 3376.22 &- &-& 3104.26  & 3198.83 &-&- \\
 		$3^{5}P_{3}$& &3648.04  &3734.96 &- &- &3459.57 &3568.53 &- &- \\
		\hline& 
	\end{tabular*}
\end{table*}

\begin{table}
	\centering
	\caption{Two meson threshold for different tetraquark states}
	\label{twomesonthreshold}
	\begin{tabular}{cccc|ccc}
		\hline	
		&\multicolumn{3}{c|}{$T_{sssq}$} &\multicolumn{3}{c}{$T_{sqqq}$}\\
		\hline
		\multirow{2}{*}{State} & \multirow{2}{15mm}{Two-meson Threshold} 
		&\multicolumn{2}{c|}{Threshold Mass}& \multirow{2}{15mm}{Two-meson Threshold} 
		&\multicolumn{2}{c}{Threshold Mass}\\
		&&Semi-relativistic&Non-relativistic&&Semi-relativistic&Non-relativistic\\   
		\hline
		$^{1}S_{0}$ & $\eta_{s}(1S) \; K_{0}(1S)$ &1258 &1258 &$\pi(1S) \; K_{0}(1S)$ &636&635\\
		\hline
		\multirow{2}{*}{$^{3}S_{1}$} & $\eta_{s}(1S) \; K_{1}(1S)$ &1660&1634& $\pi(1S) \; K_{1}(1S)$ &913&912\\
			& $\phi(1S)\; K_{0}(1S)$ &1516&1515& $\rho(1S) \; K_{0}(1S)$ &1271&1270\\
		\hline
		$^{5}S_{2}$ & $\phi(1S)\; K_{1}(1S)$ &1918&1910 & $\rho(1S) \; K_{1}(1S)$&1674&1664\\
		\hline
		\multirow{2}{*}{$^{3}P_{0}$} & $\eta_{s}(1S)$  $K_{0}(1P)$ &1835&1855  &$\pi(1S)$  $K_{0}(1P)$&1231&1233 \\
		 & $K_{0}(1S)$  $f_{0}(1P)$ &1751&1746& $K_{0}(1S)$  $a_{0}(1P)$&1358&1399 \\
		\hline
		\multirow{2}{*}{$^{3}P_{1}$} & $\eta_{s}(1S)$  $K_{1}(1P)$ &2181&2179 & $\pi(1S)$  $K_{1}(1P)$&1577&1557 \\
		 & $K_{0}(1S)$  $f_{1}(1P)$ &1950&1955 & $K_{0}(1S)$  $a_{1}(1P)$&1812&1795\\
		\hline
		\multirow{2}{*}{$^{3}P_{2}$} & $\eta_{s}(1S)$  $K_{2}(1P)$ &2234&2215 & $\pi(1S)$  $K_{2}(1P)$&1630&1593\\
		& $K_{0}(1S)$  $f_{2}(1P)$ &1984&1988 & $K_{0}(1S)$  $a_{2}(1P)$&1867&1830\\
		\hline
		\multirow{2}{*}{$^{5}P_{1}$} & $\eta_{s}(1S)$  $K_{1}(1P)$ &2134&2144 & $\pi(1S)$  $K_{1}(1P)$&1530&1522\\
		& $K_{0}(1S)$  $h_{1}(1P)$ &1777&1759 & $K_{0}(1S)$  $b_{1}(1P)$&1929&1933\\
		\hline
		\multirow{2}{*}{$^{5}P_{2}$} & $\phi(1S)$  $K_{1}(1P)$ &2457&2438 & $\rho(1S)$  $K_{1}(1P)$&2229	&2231\\
		& $K_{1}(1S)$  $f_{1}(1P)$ &2353&2349 & $K_{1}(1S)$  $a_{1}(1P)$ &2215&2190\\
		\hline
		\multirow{2}{*}{$^{5}P_{3}$} & $\phi(1S)$  $K_{2}(1P)$ &2510&2474 & $\rho(1S)$  $K_{2}(1P)$ &2262 &2265\\
		& $K_{1}(1S)$  $ f_{2}(1P)$ &2386 &2383 & $K_{1}(1S)$  $a_{2}(1P)$ &2269&2225\\
		\hline
	\end{tabular}
\end{table}

\begin{table*}
	\centering
	\caption{Tetraquark assignment for various resonances }
	\label{Assignment}
	\begin{tabular}{cccccc}
		\hline	
		State &Tetraquark &Resonance &Mass$_{\text{Cal.}}$  &Mass$_{\text{Exp.}}$ &$\Gamma$ \\
		\hline
		$1\; ^{1}S_{0_{\bar{\textbf{3}}-{\textbf{3}}}}$ &$T_{sqqq}$ &$K_{0}^{*}(1430)$ &1439.99 &$1425\pm50$ &$270\pm80$\\
		$1\; ^{1}S_{0_{\textbf{6}-\bar{{\textbf{6}}}}}$ &$T_{sqqq}$ &$K(1630)$ &$1631.54$ &$1629\pm7$ &$16^{+19}_{-16}$\\
		$1\; ^{3}S_{1_{\bar{\textbf{3}}-{\textbf{3}}}}$ &$T_{sqqq}$ &$K_{1}(1650)$ &1653.02 &$1650\pm50$ &$150\pm50$\\
		$1\; ^{5}S_{2_{\bar{\textbf{3}}-{\textbf{3}}}}$ &$T_{sqqq}$ &$K_{2}^{*}(1980)$ &2070.76 &$1990^{+60}_{-50}$ &$348^{+50}_{-30}$\\
		$2\; ^{1}S_{0_{\textbf{6}-\bar{{\textbf{6}}}}}$ &$T_{sqqq}$ &$K_{0}^{*}(2600)$ &2595.88 &$2662\pm59\pm201$ &$480\pm47\pm72$\\
		$1\; ^{1}S_{0_{\bar{\textbf{3}}-{\textbf{3}}}}$ &$T_{sssq}$ &$X(1855)$ &1845.90 &$1856.6\pm5$ &$20\pm5$\\
		$1\; ^{3}S_{1_{\bar{\textbf{3}}-{\textbf{3}}}}$ &$T_{sssq}$ &$X(1855)$ &1871.25 &$1856.6\pm5$ &$20\pm5$\\
		$2\; ^{1}S_{0_{\bar{\textbf{3}}-{\textbf{3}}}}$ &$T_{sssq}$ &$K_{0}^{*}(2600)$ &2705.68 &$2662\pm59\pm201$ &$480\pm47\pm72$\\
		
		\hline& 
	\end{tabular}
\end{table*}
	\section{Decay}
\label{sec:3}
The fundamental property following mass spectra for bound states is decay width. Considering the fact that a tetraquark is a bound state of four quarks, calculating its decay becomes as complex as its mass spectra. In the present study, decay is studied with the help of a quark rearrangement technique for hadronic decay channels. Additionally, the strong decay of tetraquark in two mesons by fall apart mechanism is discussed.

\subsection{Re-arrangement Decay}

The recoupling of spin wave functions is given by:
\begin{subequations}
	\begin{equation}		
	\bigr|{\{{(ss)}^{1}{(\bar{s}\bar{q})}^{1}\}^{0}}\bigr \rangle=-\frac{1}{2} \bigr|\{\bigr(s\bar{s}\bigr)^{1}\bigr(s\bar{q}\bigr)^{1}\bigr \}^{0}\rangle  + \frac{\sqrt{3}}{2}\bigr|\{\bigr(s\bar{s}\bigr)^{0}\bigr(s\bar{q}\bigr)^{0}\bigr\}^{0} 	 \rangle 
	\label{spinwfsssqJ0}		
\end{equation}
\begin{equation}		
	\bigr|{\{{(sq)}^{1}{(\bar{q}\bar{q})}^{1}\}^{0}}\bigr \rangle=-\frac{1}{2} \bigr|\{\bigr(s\bar{q}\bigr)^{1}\bigr(q\bar{q}\bigr)^{1}\bigr \}^{0}\rangle  + \frac{\sqrt{3}}{2}\bigr|\{\bigr(s\bar{q}\bigr)^{0}\bigr(q\bar{q}\bigr)^{0}\bigr\}^{0} \rangle  	
	\label{spinwfsqqqJ0}		
\end{equation}
\begin{equation}		
	\bigr|{\{{(ss)}^{1}{(\bar{s}\bar{q})}^{1}\}^{1}}\bigr \rangle=\frac{1}{\sqrt{2}} \bigr|\{\bigr(s\bar{s}\bigr)^{1}\bigr(s\bar{q}\bigr)^{0}\bigr\}^{1} \rangle  + \frac{{1}}{\sqrt{2}}\bigr|\{\bigr(s\bar{s}\bigr)^{0}\bigr(s\bar{q}\bigr)^{1}\}^{1}\bigr \rangle  	
	\label{spinwfsssqJ1}		
\end{equation}
\begin{equation}		
	\bigr|{\{{(sq)}^{1}{(\bar{q}\bar{q})}^{1}\}^{1}}\bigr \rangle=\frac{1}{\sqrt{2}} \bigr|\{\bigr(s\bar{q}\bigr)^{1}\bigr(q\bar{q}\bigr)^{0}\bigr\}^{1} \rangle  + \frac{{1}}{\sqrt{2}}\bigr|\{\bigr(s\bar{q}\bigr)^{0}\bigr(q\bar{q}\bigr)^{1}\}^{1}\bigr \rangle  	
	\label{spinwfsqqqJ1}		
\end{equation}
\begin{equation}		
	\bigr|{\{{(ss)}^{1}{(\bar{s}\bar{q})}^{1}\}^{2}}\bigr \rangle = \bigr|{\{{(s\bar{s})}^{1}{(s\bar{q})}^{1}\}^{2}}\bigr \rangle
	\label{spinwfsssqJ2}		
\end{equation}
\begin{equation}		
	\bigr|{\{{(sq)}^{1}{(\bar{q}\bar{q})}^{1}\}^{2}}\bigr \rangle= \bigr|{\{{(s\bar{q})}^{1}{(q\bar{q})}^{1}\}^{2}}\bigr \rangle
	\label{spinwfsqqqJ2}		
\end{equation}
\end{subequations}

Similarly, the recoupling of color wave functions is given by:
\begin{subequations}
\begin{equation}		
	|{{(ss)}_{\bar{\textbf{3}}}{(\bar{s}\bar{q})}_{\textbf{3}}}|\rangle=\sqrt{\frac{1}{3}} |\bigr(s\bar{s}\bigr)_{\textbf{1}}\bigr(s\bar{q}\bigr)_{\textbf{1}}\rangle - \sqrt{\frac{2}{3}}|\bigr(s\bar{s}\bigr)_{\textbf{8}}\bigr(s\bar{q}\bigr)_{\textbf{8}}\rangle 
	\label{colorwf1sssq}		
\end{equation}
\begin{equation}		
	|{{(ss)}_{\textbf{6}}}{(\bar{s}\bar{q})}_{\bar{\textbf{6}}}|\rangle=\sqrt{\frac{2}{3}} |\bigr(s\bar{s}\bigr)_{\textbf{1}}\bigr(s\bar{q}\bigr)_{\textbf{1}}\rangle + \sqrt{\frac{1}{3}}|\bigr(s\bar{s}\bigr)_{\textbf{8}}\bigr(s\bar{q}\bigr)_{\textbf{8}}\rangle 
	\label{colorwf2sssq}		
\end{equation}

\begin{equation}		
	|{{(sq)}_{\bar{\textbf{3}}}{(\bar{q}\bar{q})}_{\textbf{3}}}|\rangle=\sqrt{\frac{1}{3}} |\bigr(s\bar{q}\bigr)_{\textbf{1}}\bigr(q\bar{q}\bigr)_{\textbf{1}}\rangle - \sqrt{\frac{2}{3}}|\bigr(s\bar{q}\bigr)_{\textbf{8}}\bigr(q\bar{q}\bigr)_{\textbf{8}}\rangle 
	\label{colorwf1sqqq}		
\end{equation}
\begin{equation}		
	|{{(sq)}_{\textbf{6}}}{(\bar{q}\bar{q})}_{\bar{\textbf{6}}}|\rangle=\sqrt{\frac{2}{3}} |\bigr(s\bar{q}\bigr)_{\textbf{1}}\bigr(q\bar{q}\bigr)_{\textbf{1}}\rangle + \sqrt{\frac{1}{3}}|\bigr(s\bar{q}\bigr)_{\textbf{8}}\bigr(q\bar{q}\bigr)_{\textbf{8}}\rangle 
	\label{colorwf2sqqq}		
\end{equation}
\end{subequations}

Utilizing the Fierz transformation, $s\bar{s},q\bar{q},q\bar{s}$ and $s\bar{q}$ are brought together \cite{Ali:2019roi}, and the tetraquark decays into two mesons by the spectator pair technique. 

The quark bilinears are normalized to unity.
The rearrangement decay is described as due to the decay of individual $s\bar{s}$ pair into lower mass states in the following channels:

1. The spin-0 pair in color singlet representation decays into two gluons or two photons. The two gluons are further converted and confined in the form of lighter hadrons. This conversion occurs with a rate of order of $\alpha_{s}^{2}$. 
The decays observed are:
\begin{center}
	\[
	\begin{rcases}
	\mathrm{T_{ss\bar{s}\bar{q}} \; \; / \; \; T_{sq\bar{q}\bar{q}} \rightarrow K_{0}+\text{2 gluons}}  \\
	\mathrm{T_{ss\bar{s}\bar{q}} \; \; / \; \; T_{sq\bar{q}\bar{q}} \rightarrow K_{0}+\text{2 photons} }\\
	\end{rcases}\rightarrow\mathrm{^{1}S_{0}\;  state }
	\]
	\[
	\begin{rcases}
		\mathrm{T_{ss\bar{s}\bar{q}}\; \; / \; \; T_{sq\bar{q}\bar{q}} \rightarrow K_{1}+\text{2 gluons}}  \\
		\mathrm{T_{ss\bar{s}\bar{q}}\; \; / \; \; T_{sq\bar{q}\bar{q}} \rightarrow K_{1}+\text{2 photons} }\\
		\end{rcases}\rightarrow\mathrm{^{3}S_{1}\;  state }
	\]
\end{center}
2. The spin-1 pair in color singlet representation decays into three gluons or three photons. The three gluons are further converted and confined in the form of lighter hadrons. The rate is of order $\alpha_{s}^{3}$. The decays observed are:
\begin{center}
	\[
	\begin{rcases}
	\mathrm{T_{ss\bar{s}\bar{q}}\; \; / \; \; T_{sq\bar{q}\bar{q}} \rightarrow K_{1}+\text{3 gluons}}\\
	\mathrm{T_{ss\bar{s}\bar{q}}\; \; / \; \; T_{sq\bar{q}\bar{q}} \rightarrow K_{1}+\text{3 photons}}\\
	\end{rcases}\rightarrow\mathrm{^{1}S_{0}\; state \;and \;^{5}S_{2}\; state }
	\]
	\[
	\begin{rcases}
		\mathrm{T_{ss\bar{s}\bar{q}}\; \; / \; \; T_{sq\bar{q}\bar{q}} \rightarrow K_{0}+\text{3 gluons}}\\
		\mathrm{T_{ss\bar{s}\bar{q}}\; \; / \; \; T_{sq\bar{q}\bar{q}} \rightarrow K_{0}+\text{3 photons}}\\
	\end{rcases}\rightarrow\mathrm{^{3}S_{1}\; state}
	\]
\end{center}

3. The spin-0 and spin-1 $q\bar{q}$ and $s\bar{s}$ pair in the color octet representation annihilates into one gluon and two gluons respectively. The decays observed are:
\begin{center}	
	\[
	\begin{rcases}
	\mathrm{T_{ss\bar{s}\bar{q}}\; \; / \; \; T_{sq\bar{q}\bar{q}} \rightarrow K_{0}+\text{1 gluon}} \\
	\mathrm{T_{ss\bar{s}\bar{q}}\; \; / \; \; T_{sq\bar{q}\bar{q}} \rightarrow K_{1} + \text{2 gluons}} \\
	\end{rcases}\rightarrow\mathrm{^{1}S_{0}\; state \;and \;^{5}S_{2}\; state}
	\]
	\[
	\begin{rcases}
		\mathrm{T_{ss\bar{s}\bar{q}}\; \; / \; \; T_{sq\bar{q}\bar{q}} \rightarrow K_{1}+\text{1 gluon}} \\
		\mathrm{T_{ss\bar{s}\bar{q}}\; \; / \; \; T_{sq\bar{q}\bar{q}} \rightarrow K_{0} + \text{2 gluons}} \\
	\end{rcases}\rightarrow\mathrm{^{3}S_{1}\; state}
	\]
\end{center}

4. The spin-0 $q\bar{s}$ pair in color singlet representation decays into leptonic channels $l^{+}+\nu_{l}$ when q is up quark, where $l$ is lepton ($e$ and $\mu$). Similarly, the spin-0 $s\bar{q}$ pair decays into leptonic channels $l^{-}+\nu_{l}$ . This conversion occurs with a rate of order of $\alpha^{2}$. With the spectator $q\bar{s}$ and $s\bar{q}$ pairs in account, the decays observed are:
\begin{center}
	\[
\begin{rcases}
	\mathrm{T_{ss\bar{s}\bar{q}}\rightarrow \eta_{s} + l^{\mp}\nu_{l}}\\
	\mathrm{T_{sq\bar{q}\bar{q}}\rightarrow \pi + l^{\mp}\nu_{l}}\\
	\end{rcases}\rightarrow\mathrm{^{1}S_{0}\; \; state }
	\]
		\[
	\begin{rcases}
		\mathrm{T_{ss\bar{s}\bar{q}}\rightarrow \phi + l^{\mp}\nu_{l}}\\
		\mathrm{T_{sq\bar{q}\bar{q}}\rightarrow \rho + l^{\mp}\nu_{l}}\\
	\end{rcases}\rightarrow\mathrm{^{3}S_{1}\; \; state }
	\]
	
\end{center}
5. The spin-1 $q\bar{s}$ pair in color singlet representation can also decay into the rare leptonic channels $\gamma l^{+}+\nu_{l}$ when q is up quark, where $l$ is lepton ($e$ and $\mu$). Similarly, the spin-1 $s\bar{q}$ pair decays into the rare leptonic channels $\gamma l^{-}+\nu_{l}$. This conversion occurs with a rate of order of $\alpha^{2}$. With the spectator $q\bar{s}$ and $s\bar{q}$ pairs in account, the decays observed are:

\begin{center}
	\[
	\begin{rcases}
	\mathrm{T_{ss\bar{s}\bar{q}}\rightarrow \phi + \gamma l^{\mp}\nu_{l}}\\
	\mathrm{T_{sq\bar{q}\bar{q}}\rightarrow \rho + \gamma l^{\mp}\nu_{l}}\\
	\end{rcases}\rightarrow\mathrm{^{1}S_{0}\; state \;and \;^{5}S_{2}\; state}
	\]
	\[
	\begin{rcases}
	\mathrm{T_{ss\bar{s}\bar{q}}\rightarrow \eta_{s} + \gamma l^{\mp}\nu_{l}}\\
	\mathrm{T_{sq\bar{q}\bar{q}}\rightarrow \pi + \gamma l^{\mp}\nu_{l}}\\
	\end{rcases}\rightarrow\mathrm{^{3}S_{1}\; state}
	\]
\end{center}

The ratio of overlap probabilities of the decaying meson in  tetraquark is proportional to the decay rates:
\begin{equation}
	\varrho = \frac{|\Psi_{\text{Tetraquark}}(0)|^{2}}{|\Psi_{\text{Meson}}(0)|^{2}}
\end{equation}

The individual decay rate is obtained using the formula \cite{Berestetskii:1982qgu}:
\begin{equation}
	\Gamma(\text{Meson})^{spin}_{color}=||\Psi_{T}(0)|^{2}v\sigma((\text{Meson})^{spin}_{color}\rightarrow f)
\end{equation}
where $|\Psi(0)_{T}|^{2},v,$ and $\sigma$ are the overlap probability of the annihilating pair, relative velocity, and the spin-averaged annihilation cross section in the final state $f$, respectively. The total of all the individual decay rates yields the total decay rate. A more detailed explanation of the working mechanism of the decays described in this section can be found in our previous studies, Refs. \cite{Lodha:2024yfn, Lodha:2024bwn, Lodha:2024qby}.The spectator pairs $s\bar{s}$ can appear as $\eta_{s}$ or $\phi$ on the mass shell, or combine with the outgoing $q\bar{q}$ into a pair of open-strange particles for the given tetraquark. Similarly, the spectator pair $q\bar{q}$ can appear as $\rho$ or $\pi$ on the mass shell and the spectator pair $s\bar{q}$ can appear as $K_{0}$ or $K_{1}$ on the mass shell, respectively. 
The results for calculated the decay widths for the $^{1}S_{0}$, $^{3}S_{1}$ and $^{5}S_{2}$ states are tabulated in table \ref{spectatordecay}, \ref{spectatordecayS1} and \ref{spectatordecayS2} respectively.

\begin{table*}
	\centering
	\caption{Decay width for various decay channels for $T_{^{1}S_{0}}$ in MeV}
	\label{spectatordecay}
	\begin{tabular}{cccccc}
		\hline
		\multirow{2}{*}{Tetraquark} &\multirow{2}{*}{Decay Channel}  &\multicolumn{2}{c}{$\bar{\textbf{3}}-\textbf{3}$} &\multicolumn{2}{c}{$\textbf{6}-\bar{\textbf{6}}$} \\
		&& Semi-Relativistic & Non-Relativistic & Semi-Relativistic & Non-Relativistic \\
		\hline
		&&&&&\\
		\multirow{10}{*}{$T_{sssq}$} &$K_{0}+\text{ 2 gluons}$ &0.5426 &0.5346&3.237&3.091 \\
		&$K_{0}+\text{ 2 photons}$ &$2.297\times10^{-3}$ &$2.264\times10^{-3}$&$13.707\times10^{-3}$&$13.086\times10^{-3}$\\
		&$K_{1}+\text{ 3 gluons}$  &4.087 &4.027&24.384&23.279\\
		&$K_{1}+\text{ 3 photons}$   &$3.419\times10^{-6}$ &$3.369\times10^{-6}$&$20.399\times10^{-6}$&$19.475\times10^{-6}$\\
		&$K_{0}+\text{ 1 gluons}$  &4.237 &4.175&6.319&6.033\\
		&$K_{1}+\text{ 2 gluons}$  &0.314 &0.309&0.468&0.447\\
		&$\eta_{s}+ e^{\mp}\nu_{e}$  &$1.229\times10^{-3}$ &$1.816\times10^{-3}$&$7.324\times10^{-3}$&$10.473\times10^{-3}$\\
		&$\eta_{s}+ \mu^{\mp}\nu_{\mu}$  &49.393 &72.999&294.241&420.733\\
		&$\phi+\gamma e^{\mp}\nu_{e}$  &$0.148\times10^{-3}$ &$0.218\times10^{-3}$&$0.879\times10^{-3}$&$1.257\times10^{-3}$\\
		&$\phi+\gamma \mu^{\mp}\nu_{\mu}$   &0.049 &0.073&0.297&$0.425$\\
		&&&&&\\
		
		\hline
		&&&&&\\
		
		\multirow{10}{*}{$T_{sqqq}$} &$K_{0}+\text{ 2 gluons}$ &$16.559\times10^{-3}$ &$25.589\times10^{-3}$&$110.561\times10^{-3}$&$169.151\times10^{-3}$\\
		&$K_{0}+\text{ 2 photons}$  
		&$0.426\times10^{-6}$ &$0.691\times10^{-6}$&$2.845\times10^{-6}$&$2.856\times10^{-6}$\\
		&$K_{1}+\text{ 3 gluons}$  &147.137 &238.80&982.376&1502.972\\
		&$K_{1}+\text{ 3 photons}$ 
		&$0.053\times10^{-9}$ &$0.034\times10^{-9}$&$0.358\times10^{-9}$&$0.359\times10^{-9}$\\
		&$K_{0}+\text{ 1 gluons}$  &$0.094$ &$0.086$&$0.158$&$0.158$\\
		&$K_{1}+\text{ 2 gluons}$  &$7.007\times10^{-3}$ &$6.425\times10^{-3}$&$11.696\times10^{-3}$&$11.738\times10^{-3}$\\
		&$\pi+ e^{\mp}\nu_{e}$  &$0.974\times10^{-3}$ &$1.502\times10^{-3}$&$8.515\times10^{-3}$&$13.028\times10^{-3}$\\
		&$\pi+ \mu^{\mp}\nu_{\mu}$  &39.144 &60.345&342.12&523.42\\
		&$\rho+\gamma e^{\mp}\nu_{e}$  &$0.117\times10^{-3}$ &$0.180\times10^{-3}$&$1.022\times10^{-3}$&$1.564\times10^{-3}$\\
		&$\rho+\gamma \mu^{\mp}\nu_{\mu}$   &0.039 &0.061&0.346&0.530\\
		 & &&&&\\
		\hline
	\end{tabular}
\end{table*}

\begin{table*}
	\centering
	\caption{Decay width for various decay channels for $T_{^{3}S_{1}}$ in MeV}
	\label{spectatordecayS1}
	\begin{tabular}{cccc}
		\hline
		{Tetraquark} &{Decay Channel}  	& Semi-Relativistic & Non-Relativistic  \\
		\hline
		&&&\\
		\multirow{10}{*}{$T_{sssq}$} &$K_{1}+\text{ 2 gluons}$ &1.085&1.069\\
		&$K_{1}+\text{ 2 photons}$&$4.595\times10^{-3}$&$4.529\times10^{-3}$\\
		&$K_{0}+\text{ 3 gluons}$  &2.725&2.685\\
		&$K_{0}+\text{ 3 photons}$ &$2.279\times10^{-6}$&$2.246\times10^{-6}$\\
		&$K_{1}+\text{ 1 gluons}$  &8.474&8.351\\
		&$K_{0}+\text{ 2 gluons}$  &0.209&0.206\\
		&$\phi+ e^{\mp}\nu_{e}$ &$2.458\times10^{-3}$&$3.632\times10^{-3}$\\
		&$\phi+ \mu^{\mp}\nu_{\mu}$&98.787&146.011\\
		&$\eta_{s}+\gamma e^{\mp}\nu_{e}$ &$0.099\times10^{-3}$&$0.145\times10^{-3}$\\
		&$\eta_{s}+\gamma \mu^{\mp}\nu_{\mu}$   &0.033&0.049\\
		&&&\\
		
		\hline
		&&&\\
		
		\multirow{10}{*}{$T_{sqqq}$} &$K_{1}+\text{ 2 gluons}$&$33.118\times10^{-3}$&$49.677\times10^{-3}$\\
		&$K_{1}+\text{ 2 photons}$  &$0.852\times10^{-6}$&$1.382\times10^{-6}$\\
		&$K_{0}+\text{ 3 gluons}$ &98.091&159.199\\
		&$K_{0}+\text{ 3 photons}$ &$0.035\times10^{-9}$&$0.023\times10^{-9}$\\
		&$K_{1}+\text{ 1 gluons}$ &0.188&0.172\\
		&$K_{0}+\text{ 2 gluons}$ &$4.671\times10^{-3}$&$4.283\times10^{-3}$\\
		&$\rho+ e^{\mp}\nu_{e}$ &$1.948\times10^{-3}$&$3.004\times10^{-3}$\\
		&$\rho+ \mu^{\mp}\nu_{\mu}$ &78.288&120.69\\
		&$\pi+\gamma e^{\mp}\nu_{e}$ &$0.078\times10^{-3}$&$0.120\times10^{-3}$\\
		&$\pi+\gamma \mu^{\mp}\nu_{\mu}$   &0.026&0.041\\
		& &&\\
		\hline
	\end{tabular}
\end{table*}

\begin{table*}
	\centering
	\caption{Decay width for various decay channels for $T_{^{5}S_{2}}$ in MeV}
	\label{spectatordecayS2}
	\begin{tabular}{cccc}
		\hline
		{Tetraquark} &{Decay Channel}  		& Semi-Relativistic & Non-Relativistic   \\
		\hline
		&&&\\
		\multirow{6}{*}{$T_{sqqq}$}		&$K_{1}+\text{ 3 gluons}$  &5.449 &5.369\\
		&$K_{1}+\text{ 3 photons}$ &$4.558\times10^{-6}$ &$4.492\times10^{-6}$\\
		&$K_{0}+\text{ 1 gluons}$   &5.649&5.567\\
		&$K_{1}+\text{ 2 gluons}$  &0.418&0.412\\
		&$\phi+\gamma e^{\mp}\nu_{e}$ &$0.197\times10^{-3}$&$0.218\times10^{-3}$\\
		&$\phi+\gamma \mu^{\mp}\nu_{\mu}$   &0.065&0.097\\
		&&&\\
		
		\hline
		&&&\\
		
		\multirow{6}{*}{$T_{sqqq}$} &$K_{1}+\text{ 3 gluons}$ &196.182&318.403\\
		&$K_{1}+\text{ 3 photons}$ &$0.070\times10^{-9}$&$0.045\times10^{-9}$\\
		&$K_{0}+\text{ 1 gluons}$ &0.125&0.115\\
		&$K_{1}+\text{ 2 gluons}$ &$9.342\times10^{-3}$&$8.566\times10^{-3}$\\
		&$\rho+\gamma e^{\mp}\nu_{e}$ &$0.156\times10^{-3}$&$0.241\times10^{-3}$\\
		&$\rho+\gamma \mu^{\mp}\nu_{\mu}$  &0.052&0.081\\
		& &&\\
		\hline
	\end{tabular}
\end{table*}

\subsection{Strong Decay}
	A tetraquark with diquark antidiquark formalism can decay strongly by fall fall-apart mechanism into respective mesons. The partial decay width of this process for various channels is given by,
\begin{subequations}
\begin{equation}
	\Gamma_{{T}\rightarrow \; V_{1}V_{2}}= \mathcal{G}_{T \; V_{1}V_{2}}^{2}  \frac{\lambda(T, \; V_{1},V_{2})}{8\pi}\biggr(\frac{m_{V_{1}}^{2}{m_{V_{2}}^{2}}}{m_{T}^{2}}+\frac{2\lambda^{2}(T,V_{1},V_{2})}{3} \biggr)
\end{equation} 
\begin{equation}
	\Gamma_{T\rightarrow P_{1}P_{2}}= \mathcal{G}_{T P_{1}P_{2}}^{2}  \frac{\lambda(T,P_{1},P_{2})}{8\pi}\bigr(m_{P_{1}}m_{P_{2}} + \lambda^{2}(T,P_{1},P_{2})  \bigr)
\end{equation} 
\end{subequations}
\begin{equation}
	\lambda(m_{1},m_{2},m_{3}) = \frac{\sqrt{m_{1}^{4}+m_{2}^{4}+m_{3}^{4}-2(m_{1}^{2}m_{2}^{2}+m_{2}^{2}m_{3}^{2}+m_{1}^{2}m_{3}^{2})}}{2m_{1}}
\end{equation}
	where $\mathcal{G}_{T P_{1}P_{2}}^{2} $ and $\mathcal{G}_{TV_{1}V_{2}}$ are the coupling constants for the process $T\rightarrow P_{1}P_{2}$ and $T\rightarrow V_{1}V_{2}$, while $V_{1},V_{2},P_{1},P_{2}$ are decaying vector and pseudoscalar mesons respectively. 
\begin{center}
	\begin{table}[H]
		\centering
		\caption{Partial decay width for strong decay for varioud decay channels of strange tetraquark in semi-relativistic and non-relativistic formalism }
		\label{decay}
		\begin{tabular}{cccccccc}
			\hline
			Tetraquark &State &Color Configuration & Decay Channel & $\lambda_{SR}(GeV)$  &$\lambda_{NR}(GeV)$  &$\Gamma_{SR}(GeV)$  &$\Gamma_{NR}(GeV)$  \\
			\hline
			\multirow{2}{*}{$T_{sssq}$}&$^{1}S_{0}$&$\bar{\textbf{3}}-\textbf{3}$&$\eta K_{0}$ &0.567 &0.668 &$0.0155\mathcal{G}_{T \eta_{s}K_{0}}^{2} $  &$0.0219\mathcal{G}_{T \eta_{s}K_{0}}^{2} $  \\
			&$^{5}S_{2}$&$\bar{\textbf{3}}-\textbf{3}$&$\phi K_{1}$ &0.667 &0.640 &$0.0119\mathcal{G}_{T \phi K_{1}}^{2} $  &$0.0109\mathcal{G}_{T \phi K_{1}}^{2} $  \\			\hline
			\multirow{3}{*}{$T_{sqqq}$}&$^{1}S_{0}$&$\bar{\textbf{3}}-\textbf{3}$&$\pi K_{0}$ &0.625 &0.509 &$0.0115\mathcal{G}_{T \pi K_{0}}^{2} $  &$0.0067\mathcal{G}_{T \pi K_{0}}^{2} $  \\
			&$^{1}S_{0}$&${\textbf{6}}-\textbf{\textbf{6}}$&$\pi K_{0}$ &0.733 &0.166 &$0.0176\mathcal{G}_{T \pi K_{0}}^{2} $  &$0.0006\mathcal{G}_{T \pi K_{0}}^{2} $  \\
			&$^{5}S_{2}$&$\bar{\textbf{3}}-\textbf{3}$&$\rho K_{1}$ &0.609 &0.673 &$0.0087\mathcal{G}_{T \rho K_{1}}^{2} $  &$0.0108\mathcal{G}_{T \rho K_{1}}^{2} $  \\
						\hline
		\end{tabular}	
	\end{table}
\end{center}

	\section{Regge trajectories}
\label{sec:4}
\begin{figure*}[t]
	\centering
	\begin{subfigure}{0.45\textwidth}
		\includegraphics[width=1\linewidth, height=0.3\textheight]{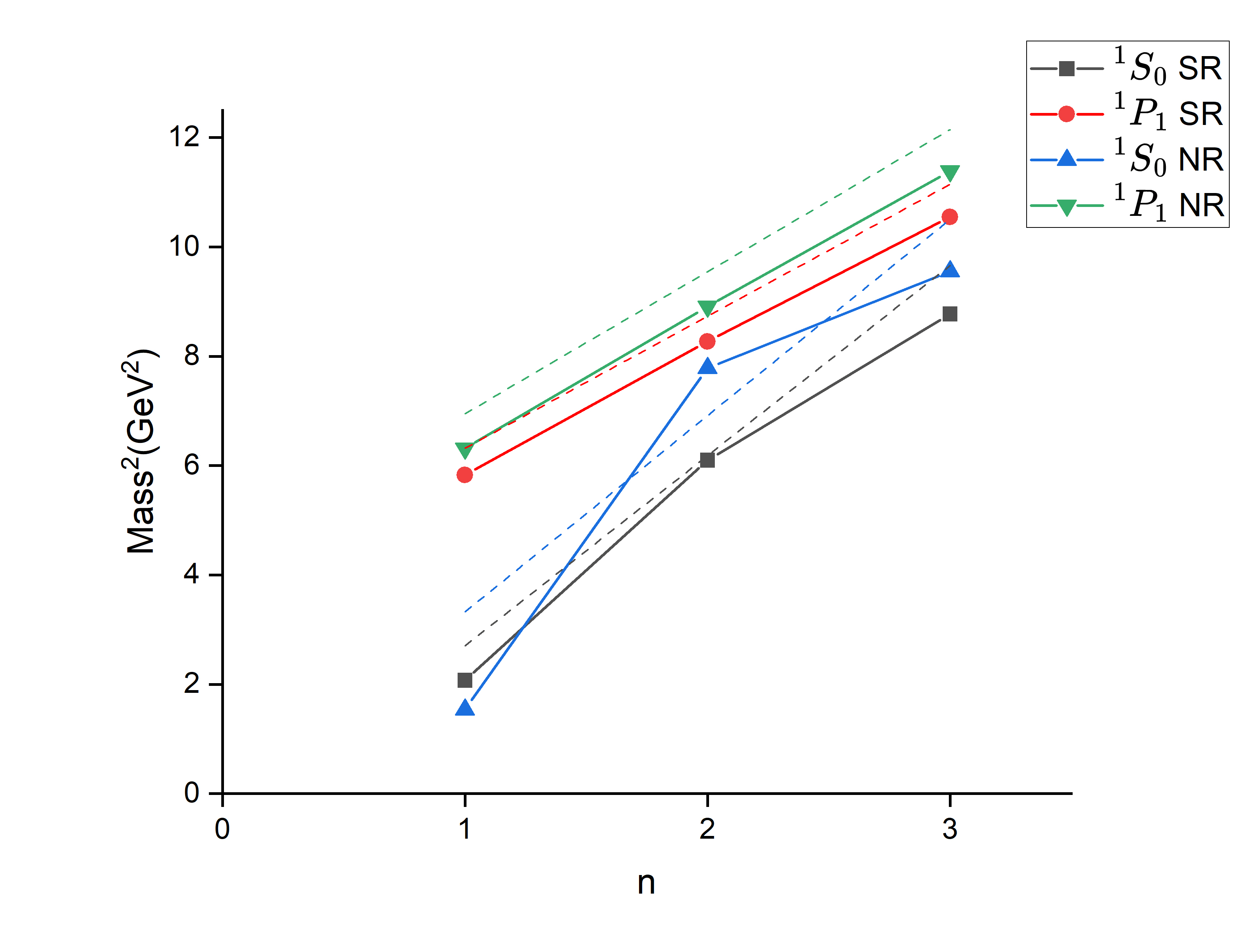}
		\caption{$T_{sqqq}$}
		\label{fig:mesonnrgraph1nat}
	\end{subfigure}
	\begin{subfigure}{0.45\textwidth}
		\includegraphics[width=1\linewidth, height=0.3\textheight]{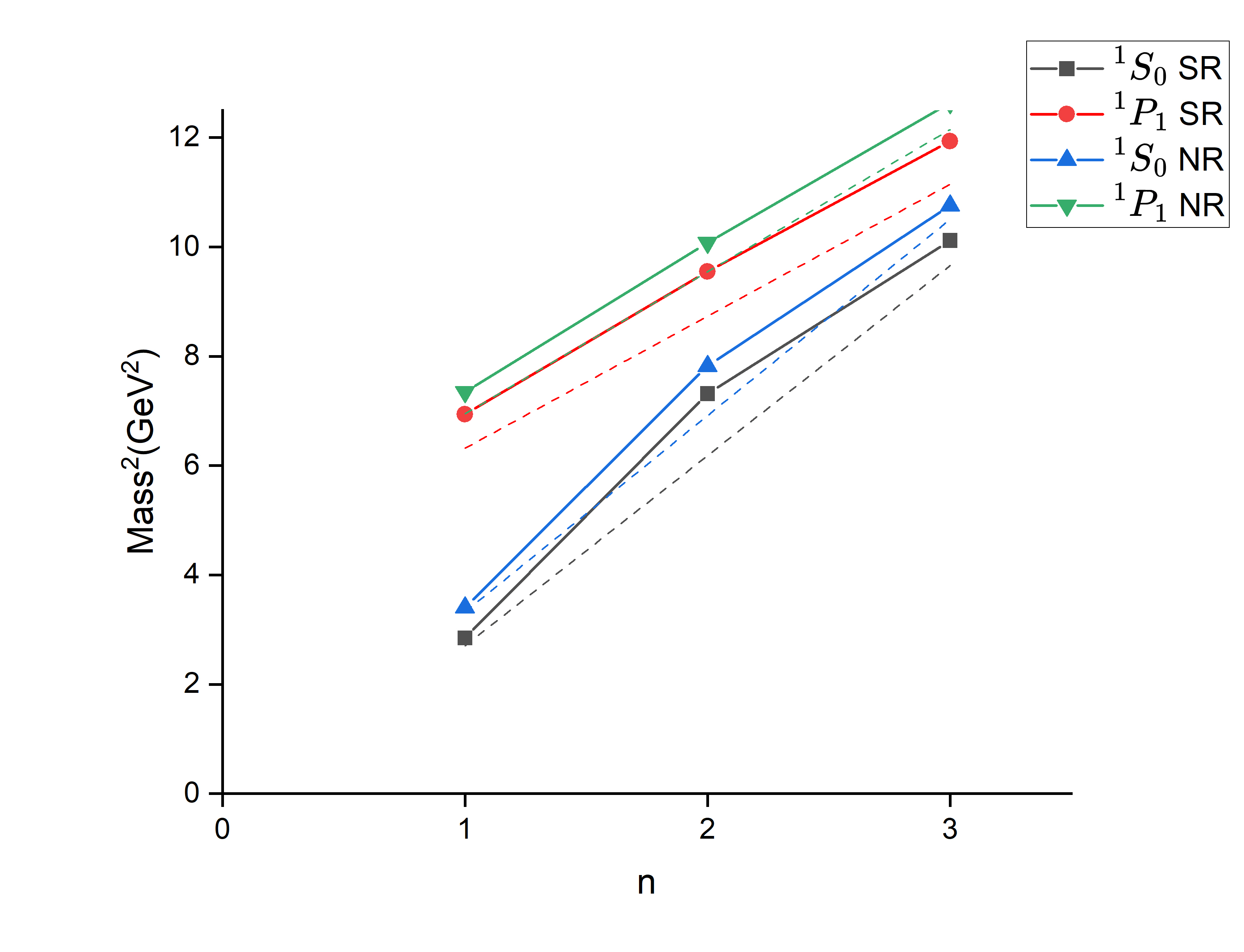}
		\caption{$T_{sssq}$}
		\label{fig:mesonsrgraph1nat}
	\end{subfigure}
	\label{GraphS0}
	\caption[]{Regge trajectory in the $(n, M^{2})$ plane for tetraquarks with Spins S = 0}
\end{figure*}

\begin{figure*}[t]
	\centering
	\begin{subfigure}{0.475\textwidth}
		\includegraphics[width=1\linewidth, height=0.3\textheight]{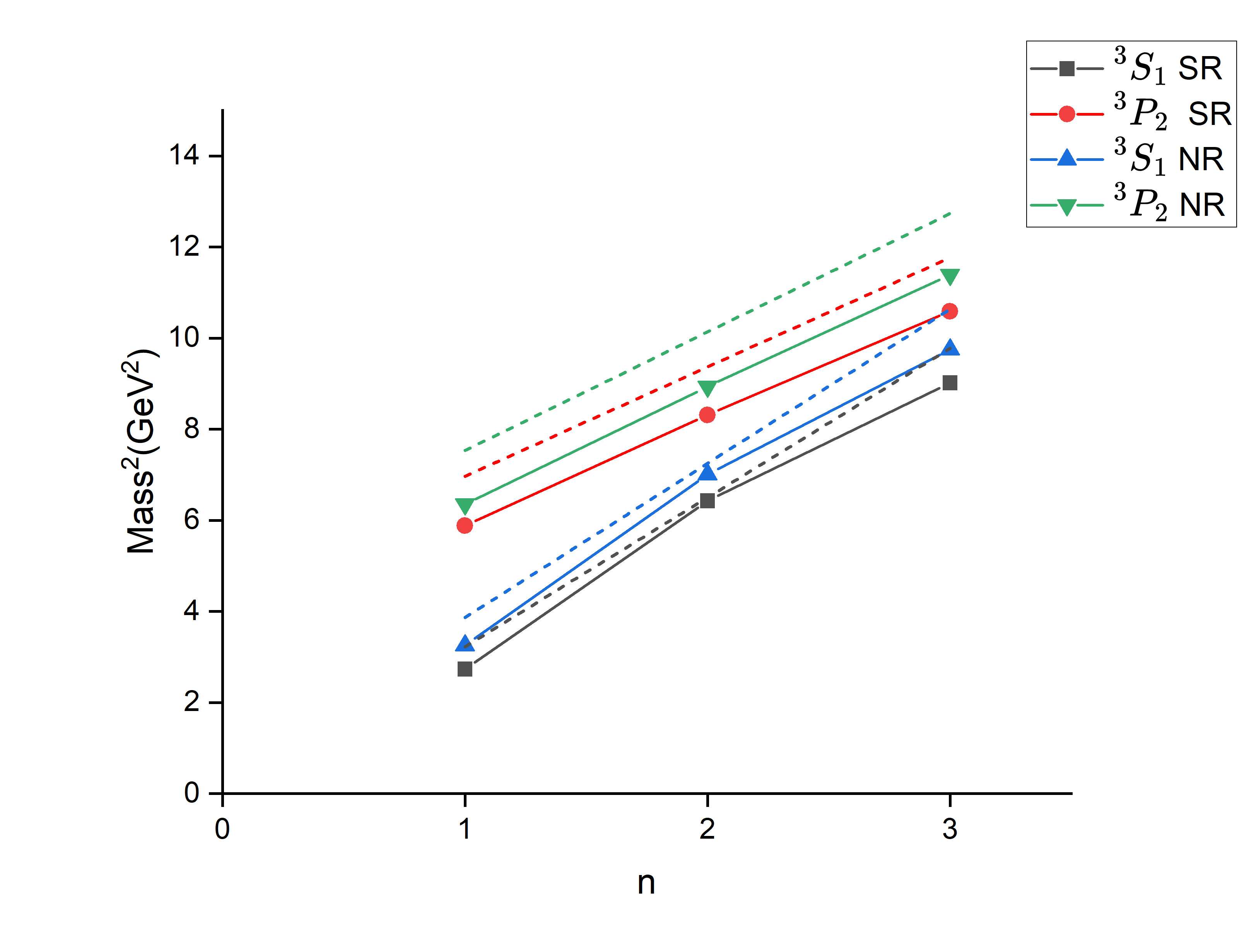}
		\caption{$T_{sqqq}$}
		\label{fig:mesonnrgraph2nat}
	\end{subfigure}
	\begin{subfigure}{0.475\textwidth}
		\includegraphics[width=1\linewidth, height=0.3\textheight]{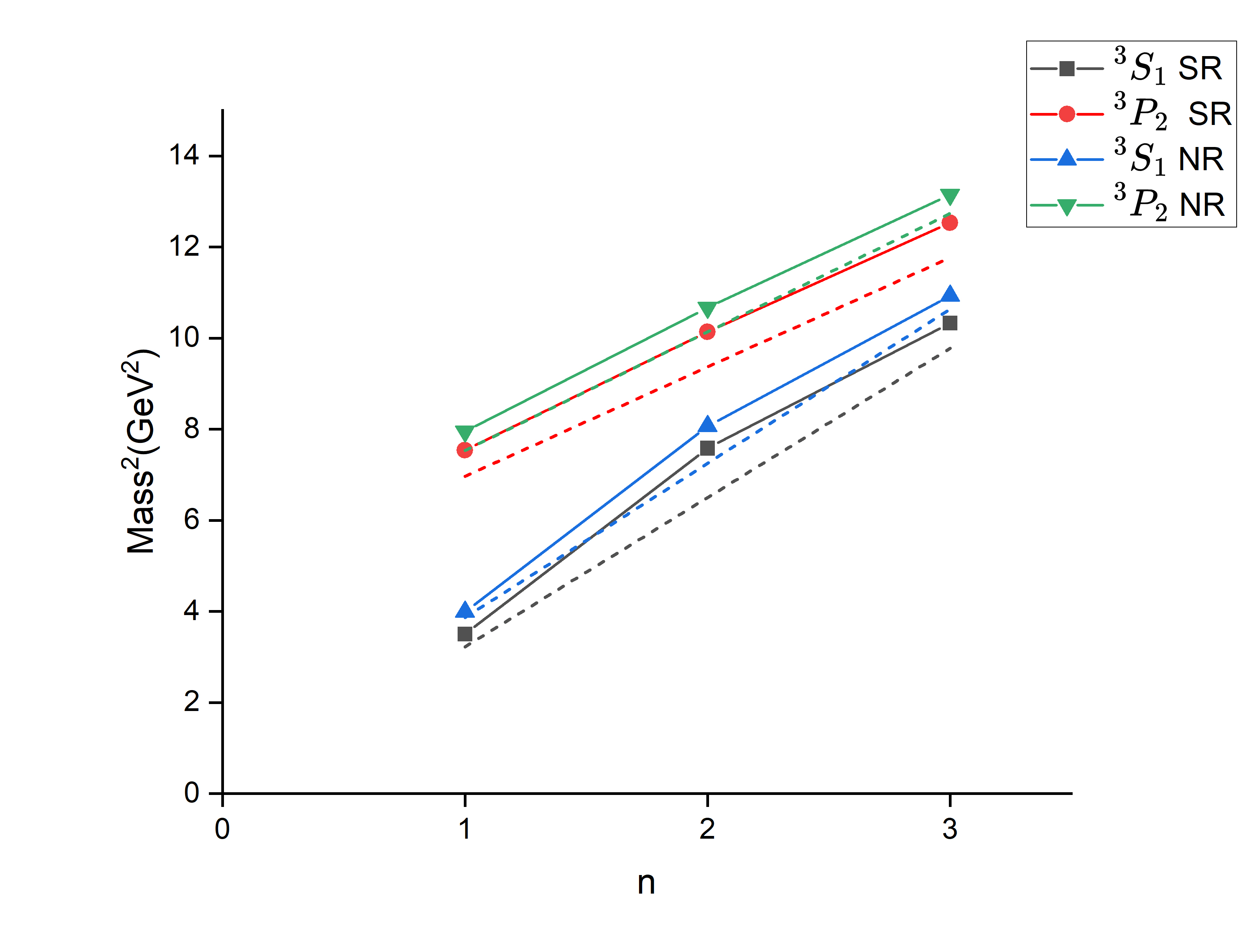}
		\caption{$T_{sssq}$}
		\label{fig:mesonsrgraph2nat}
	\end{subfigure}
	\label{GraphS1}
	\caption[]{Regge trajectory in the $(n, M^{2})$ plane for tetraquarks with Spins S = 1}
\end{figure*}

\begin{figure*}[t]
	\centering
	\begin{subfigure}{0.475\textwidth}
		\includegraphics[width=1\linewidth, height=0.3\textheight]{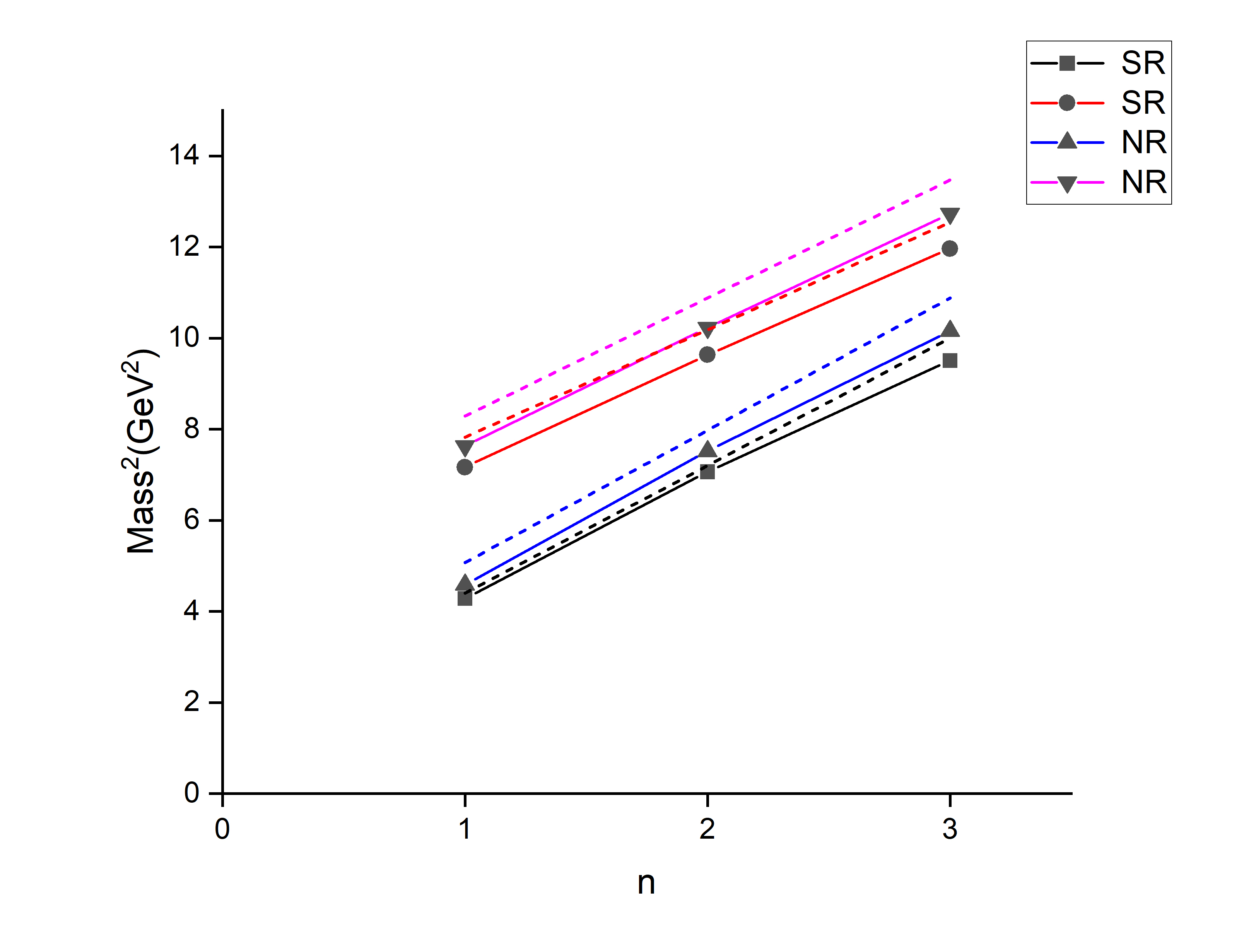}
		\caption{$T_{sqqq}$}
		\label{fig:tetraS0}
	\end{subfigure}
	\begin{subfigure}{0.475\textwidth}
		\includegraphics[width=1\linewidth, height=0.3\textheight]{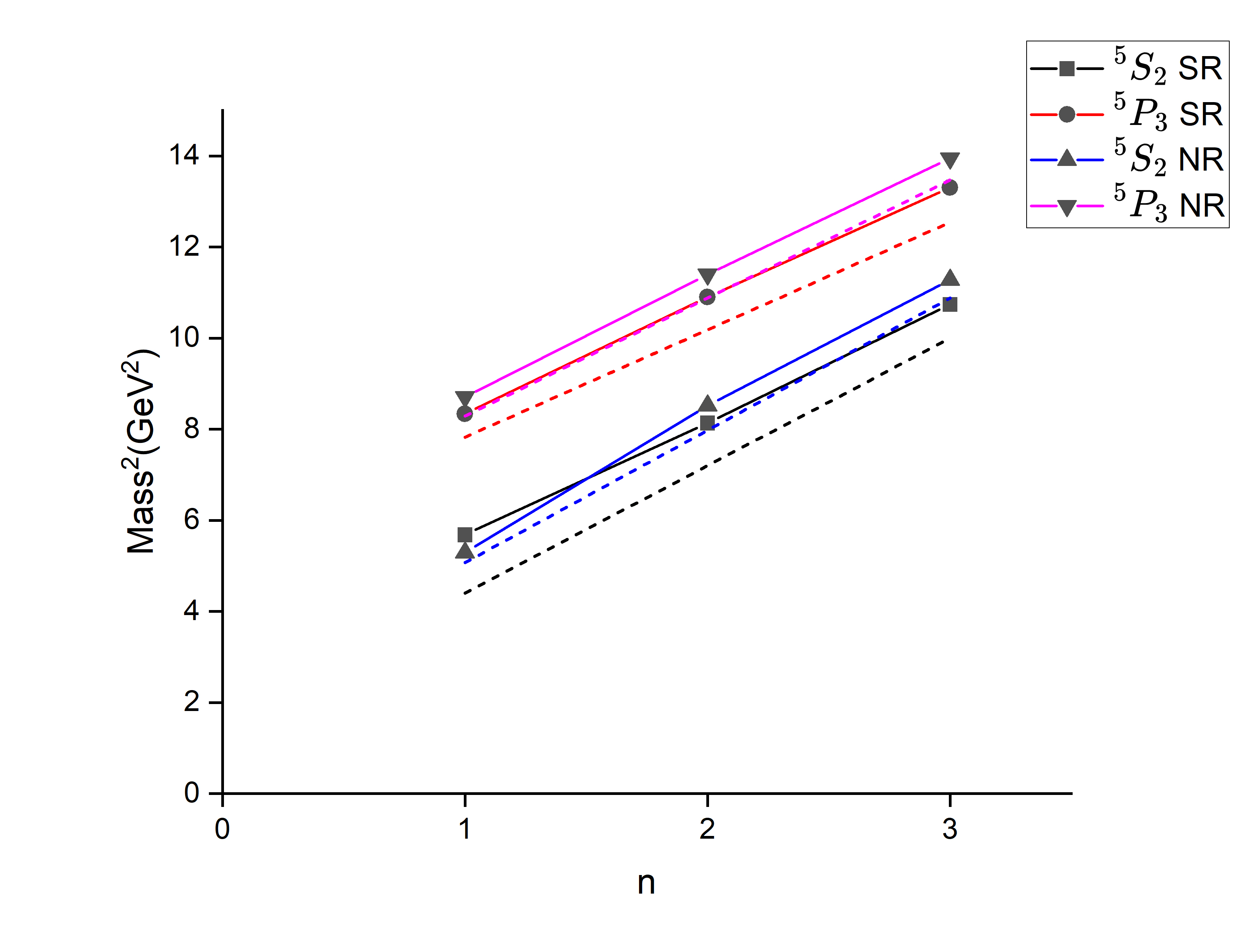}
		\caption{$T_{sssq}$}
		\label{fig:tetraS1}
	\end{subfigure}
	\label{GraphS2}
	\caption[]{Regge trajectory in the $(n, M^{2})$ plane for tetraquarks with Spins S = 2}
\end{figure*}

\vspace{30mm}
Regge trajectories play an important role in the understanding physics of bound states, particularly in the context of hadrons. The concept of these trajectories arose from the Regge theory, which was developed to explain the behavior of high-energy scattering amplitudes, but later proved to be a powerful tool in understanding the properties and classification of hadrons. This section discusses regge trajectories of the calculated mass spectrum tetraquarks. An in-depth discussion on regge trajectories and their behavior for various bound states has been done in ref. \cite{Lodha:2024yfn}. The Regge trajectories in the $(n,M^{2})$ plane are plots of the principal quantum number, n, against the square of the resonance mass, $M^{2}$. The Regge trajectories in the $(n, M^{2})$ plane are drawn for the tetraquarks for various states with S = 0, 1 and 2. as shown in Figs.  \ref{fig:mesonnrgraph1nat},  \ref{fig:mesonsrgraph1nat}, \ref{fig:mesonnrgraph2nat}, \ref{fig:mesonsrgraph2nat}, \ref{fig:tetraS0} and \ref{fig:tetraS1}, respectively.

\section{Results and Discussion}
\label{sec:5}
In the present work, we have calculated the mass spectra of the $ss\bar{s}\bar{q}$ and $sq\bar{q}\bar{q}$ tetraquarks, which is tabulated in table \ref{Swavetriplet}. These masses are calculated utilizing the fitting parameters obtained from kaonic and pionic mass spectra, adopted from ref \cite{Lodha:2024yfn,Lodha:2024bwn,Lodha:2024qby} for the non-relativistic and semi-relativistic framework. The present work estimates the mass spectra for two color configurations of diquark-antidiquark formalism, namely $\bar{\textbf{3}}-\textbf{3}$ and $\textbf{6}-\bar{\textbf{6}}$. In addition to it, the decay properties of tetraquark in various decay channels have also been calculated for $^{1}S_{0},^{3}S_{1}$ and $^{5}S_{2}$ state, which are shown in table \ref{spectatordecay}, \ref{spectatordecayS1} and \ref{spectatordecayS2} respectively. The results for strong decay have also been calculated and are tabulated in table \ref{decay}.

Since no PDG data for strange tetraquarks is available, the two-meson threshold limit is used as a reference for comparison, which has been tabulated in table \ref{twomesonthreshold}. Table \ref{Assignment} shows the numerous kaon resonances showing reasonable resemblance with the description of calculated tetraquarks. 

\subsection{The $ss\bar{s}\bar{q}$ Tetraquark}

The present study estimates the mass for $ss\bar{s}\bar{q}$ tetraquark between 1.6 GeV to 3.7 Gev for $\bar{\textbf{3}}-\textbf{3}$ color configuration and 0.9 GeV to 4.9 GeV for $\textbf{6}-\bar{\textbf{6}}$ color configuration, making it an ideal candidate to study at facilities like J-Parc and PaNDa. The $\bar{\textbf{3}}-\textbf{3}$ color configuration has 10 states while the $\textbf{6}-\bar{\textbf{6}}$ color configuration has 5 states. The spectator decay for the $^{1}S_{0}$ state, tabulated in table \ref{spectatordecay} shows $\eta_{s}+\mu^{\mp}\nu_{\mu}$ as the major branch for decay followed by $K_{0}+\text{ 1 gluon}$ and $K_{1}+\text{ 3 gluon}$. While, the decay width for $^{1}S_{0}$ state in spectator decay for $\bar{\textbf{3}}-\textbf{3}$ color configuration is under 100 MeV for both formalisms, the same for $\textbf{6}-\bar{\textbf{6}}$ color configuration is about nearly six-fold, indicating shorter half-life of the particle in that color configuration. In case of the $^{3}S_{1}$ state, $\phi + \mu^{mp}\nu_{\mu}$ is the leading decay channel, as tabulated in table \ref{spectatordecayS1}. In case of the $^{5}S_{2}$ state, $K_{0} + \text{1gluon}$ and $K_{1} + \text{3gluon}$ are the leading decay channels, as tabulated in table \ref{spectatordecayS2}. The  $ss\bar{s}\bar{q}$ tetraquark can decay by strong decay in channels predicted in table \ref{decay}. The regge trajectories for this tetraquark are also well-behaved. Except for the ground state of $^{1}S_{0}$ in $\textbf{6}-\bar{\textbf{6}}$ color configuration, all the calculated states for this tetraquark in the present study lie above the two meson threshold calculated in table \ref{twomesonthreshold}. Two resonances show agreeable results with the calculated mass for $ss\bar{s}\bar{q}$ tetraquark, namely $X(1855)$ and $K^{*}_{0}(2600)$. The $X(1855)$ resonance has an average mass of $1856.6\pm5$ MeV and undefined $J^{P}$. This resonance was observed in antiproton annihilations in Deuterium at rest into two Pions with an average decay rate of $20\pm5$ MeV \cite{Bridges:1986vc}. The mass of ground state $^{1}S_{0}$ $ss\bar{s}\bar{q}$ tetraquark in non-relativistic formalism in the present work is found to be $1845.90$ MeV and is in acceptable mass range to be considered a potential candidate for $X(1855)$. Similarly, the mass of ground state $^{3}S_{0}$ $ss\bar{s}\bar{q}$ tetraquark in semi-relativistic formalism in the present work is found to be $1871.25$ MeV and is very near to the mass of $X(1855)$ resonance. Another resonance that shows good agreement with the calculated mass of $ss\bar{s}\bar{q}$ tetraquark is $K_{0}^{*}(2600)$. The average mass of $K_{0}^{*}(2600)$ is $2662\pm59\pm201$, which is under good agreement with the $2 ^{1}S_{0}$ state of $ss\bar{s}\bar{q}$ tetraquark in semi-relativistic formalism of the present work which is $2705.68$ MeV. $K_{0}^{*}(2600)$ has been observed in the study of charmonium decays to $K_{s}^{0}K\pi$ in the $B\rightarrow(K_{s}^{0}K\pi) K$ channels with an average decay width of $480\pm47\pm72$ MeV \cite{LHCb:2023evz}. The $I(J^{P})$ value of $K_{0}^{*}(2600)$ is $1/2(0^{+})$ which is identical for $2 ^{1}S_{0}$ state of $ss\bar{s}\bar{q}$ tetraquark. All the predicted states for $ss\bar{s}\bar{q}$ tetraquark have the internal color configuration $\bar{\textbf{3}}-\textbf{3}$. 

\subsection{The $sq\bar{q}\bar{q}$ Tetraquark}

The present study estimates the mass for $sq\bar{q}\bar{q}$ tetraquark between 1.2 GeV to 3.6 Gev for $\bar{\textbf{3}}-\textbf{3}$ color configuration and 0.7 GeV to 4.8 GeV for $\textbf{6}-\bar{\textbf{6}}$ color configuration, making it an ideal candidate to study at facilities like J-Parc and PaNDa. The spectator decay for $^{1}S_{0}$ state, tabulated in table \ref{spectatordecay} indicates $K_{1}+\text{ 3 gluon}$ as the major branch for decay followed by $\pi+\mu^{\mp}\nu_{\mu}$ . While, the decay width in spectator decay for $\bar{\textbf{3}}-\textbf{3}$ color configuration is under 300 MeV for both formalisms, the same for $\textbf{6}-\bar{\textbf{6}}$ color configuration is nearly 1.5 GeV to 2.1 GeV, indicating a very short particle for that color configuration. For the $^{3}S_{1}$ state, the decay channel $K_{0}+\text{3gluons}$ emerges as the leading decay channel followed by $\rho+\mu^{\mp}+\nu_{\mu}$. The total decay width by rearrangement decay for the $^{3}S_{1}$ state is under 300 MeV.
In case of the $^{5}S_{2}$ state, the $K_{1}+\text{3 gluon}$ is the main decay channel.  The  $ss\bar{s}\bar{q}$ tetraquark can decay by strong decay in channels predicted in table \ref{decay}. The regge trajectories for this tetraquark are also well-behaved.
All the calculated states for this tetraquark in the present study lie above the two meson thresholds calculated in table \ref{twomesonthreshold}. In the case of the $sq\bar{q}\bar{q}$ tetraquark, 7 resonances show good agreement with the calculated mass. The $K^{*}(1430)$ resonance has an average mass of $1425\pm50$ MeV with a decay width of $270\pm80$ MeV. The $1 ^{1}S_{0}$ state of $T_{sqqq}$ tetraquark semi-relativistic formalism has mass 1439.99 MeV which is very near to the $K^{*}(1430)$ resonance. This state also has same $I(J^{P})$ value, $1/2(0^{+})$ and has been observed in numerous processes \cite{ParticleDataGroup:2024cfk}. The $K(1630)$ resonance has been observed in the mass spectrum of $K_{s}^{0}\pi^{+}\pi^{-}$ system produced in $\pi^{-}p$ interactions at high momentum transfers with an average mass of  $1629\pm7$ MeV and a decay width of $16^{+19}_{-16}$ MeV \cite{Karnaukhov:1998qq}. This resonance is not a well-established particle and has a $I(J^{P})$ value of $1/2(?^{?})$. This resonance is a good fit for the $1 ^{1}S_{0}$ state of the $sq\bar{q}\bar{q}$ tetraquark in the semi-relativistic formalism of the present study which has a mass of 1631.54 MeV and an identical $I(J^{P})$. The $K_{1}(1650)$ resonance is a well-established particle with an average mass of $ 1650\pm50$ and a $I(J^{P})$ value of $1/2(1^{+})$. This resonance has been  observed in $13K^{+}p\rightarrow\phi K^{+}p$ process with an average decay of $150\pm50$ MeV \cite{Frame:1985ka}. This description fits very well with the $1 ^{3}S_{1}$ tetraquark state in the semi-relativistic formalism which has identical $I(J^{P})$ value and a mass of 1653.02 MeV. The $K_{2}^{*}(1980)$ resonance is also a well-established particle with an average mass of $1990^{+60}_{-50}$ MeV. This resonance has been observed in $\psi(2S)\rightarrow K^{+}K^{-}\eta$, $J/\psi\rightarrow K^{+}K^{-}\pi^{0}$ and $11K^{-}p\rightarrow\bar{K}^{0}\pi^{+}\pi^{-}n$ process with an average decay width of $348^{+50}_{-30}$ MeV \cite{BESIII:2019dme,BESIII:2019apb}. The $1 ^{5}S_{2}$ state of the $sq\bar{q}\bar{q}$ tetraquark in the semi-relativistic formalism of the present study has an average mass of 2070.76 MeV and seems a good candidate for the $K_{2}^{*}(1980)$ resonance. Lastly, the $K^{*}_{0}(2600)$, which has been discussed earlier can also be predicted for the $2 ^{1}S_{0}$ state of the $sq\bar{q}\bar{q}$ tetraquark in the semi-relativistic formalism of the present study which has an average mass of 2595.88 MeV.

\section{Conclusion}

To conclude, the mass spectra of light-strange tetraquarks in various diquark-antidiquark configurations have been calculated in a non-relativistic and semi-relativistic framework. $J^{P}$ values have been assigned accordingly to the calculated states. Concurrently, the decay of light-strange tetraquarks has been calculated using the spectator model \cite{Becchi:2020mjz} and the rearrangement model \cite{Ali:2019roi}. Regge plots for tetraquarks have been plotted for various spins and parity. Six resonances have been inspected as potential light-strange tetraquark states, $K_{0}^{*}(1430)$, $K(1630)$, $K_{1}(1650),$  $K_{2}^{*}(1980),$ $K_{0}^{*}(2600)$ and $X(1855)$.
Utilizing the formalism of the present work, tetraquarks with multiple quark flavors will be explored in the future. This investigation will encompass an examination of their characteristics such as masses, decay widths, and magnetic moments. These findings could potentially serve as valuable input and comparative data for institutions such as PANDA, J-PARC, Belle, LHCb, and others, which focus on in-depth analyses of resonances involving strange quarks \cite{PANDA:2016fbp,Singh:2016hoh,Singh:2019jug,PANDA:2018zjt}.

\section{Data Availability Statement} The datasets generated during and/or analysed during the current study are available from the corresponding author on reasonable request.

%

\end{document}